\documentclass[10pt,journal,compsoc]{IEEEtran}
\usepackage{amsmath,amssymb,amsfonts}
\usepackage{algorithm}
\usepackage{algpseudocode}
\usepackage{graphicx}
\usepackage{textcomp}
\usepackage{cite}

\usepackage{tikz}
\usetikzlibrary{matrix,calc}

\begin{document}

\title{Regularized Shallow Image Prior for Electrical Impedance Tomography}

\author{Zhe~Liu,~\IEEEmembership{Student Member,~IEEE,} Zhou~Chen,~\IEEEmembership{Student Member,~IEEE,}
        Qi Wang,~\IEEEmembership{Member,~IEEE}, Sheng~Zhang,
        and~Yunjie~Yang,~\IEEEmembership{Member,~IEEE}% <-this % stops a space
\IEEEcompsocitemizethanks{\IEEEcompsocthanksitem Zhe Liu, Zhou Chen and Yunjie Yang are with the SMART Group, Institute for Digital Communications, School of
Engineering, The University of Edinburgh, Edinburgh EH9 3JL, U.K.
(e-mail: zz.liu@ed.ac.uk; y.yang@ed.ac.uk).\protect\\
\IEEEcompsocthanksitem Qi Wang is with the School of Electronics $\&$ Information Engineering, Tiangong University, Tianjin 300387, China (e-mail: wangqitju@163.com).\protect\\
\IEEEcompsocthanksitem Sheng Zhang is with the Peking University Shenzhen Hospital, Shenzhen, China.\protect\\
% E-mail: see http://www.michaelshell.org/contact.html
% \IEEEcompsocthanksitem J. Doe and J. Doe are with Anonymous University.
}%

\thanks{Manuscript received xxxx, xxxx; revised xxxx, xxxx.}}

\markboth{Journal of \LaTeX\ Class Files,~Vol.~14, No.~8, xxxx~xxxx}%
{Shell \MakeLowercase{\textit{et al.}}: Bare Advanced Demo of IEEEtran.cls for IEEE Computer Society Journals}

\IEEEtitleabstractindextext{
\begin{abstract}
Untrained Neural Network Prior (UNNP) based algorithms have gained increasing popularity in tomographic imaging, as they offer superior performance compared to hand-crafted priors and do not require training. UNNP-based methods usually rely on deep architectures which are known for their excellent feature extraction ability compared to shallow ones. Contrary to common UNNP-based approaches, we propose a regularized shallow image prior method that combines UNNP with hand-crafted prior for Electrical Impedance Tomography (EIT). Our approach employs a 3-layer Multi-Layer Perceptron (MLP) as the UNNP in regularizing 2D and 3D EIT inversion. We demonstrate the influence of two typical hand-crafted regularizations when representing the conductivity distribution with shallow MLPs. We show considerably improved EIT image quality compared to conventional regularization algorithms, especially in structure preservation. The results suggest that combining the shallow image prior and the hand-crafted regularization can achieve similar performance to the Deep Image Prior (DIP) but with less architectural dependency and complexity of the neural network.
% Untrained neural network prior (UNNP) based algorithms have been increasingly attractive in tomographic imaging, due to no need for training and superior performance than hand-crafted priors. UNNP-based methods usually adopt deep architectures since deep neural networks often demonstrate superior feature extraction ability compared with shallow ones. Contrary to common UNNP-based approaches, we propose a regularized shallow image prior approach that combines UNNP with a hand-crafted smoothing prior for electrical impedance tomography (EIT). Our approach employs a 3-layer multi-layer perceptrons (MLP) as effective models in regularizing 2D and 3D EIT inversion. We demonstrate the influence of two typical regularisers when regularizing the conductivity distribution with shallow MLPs. Compared with typical regularisation algorithms, we show considerably improved EIT image quality, especially in structure preservation. The results suggest that the combination of the shallow image prior and regularisation could achieve similar performance with the deep image prior but with less computational cost.
\end{abstract}

}

\maketitle
\IEEEdisplaynontitleabstractindextext
\IEEEpeerreviewmaketitle

\ifCLASSOPTIONcompsoc
\IEEEraisesectionheading{\section{Introduction}\label{sec:introduction}}
\else

\section{Introduction}
\label{sec:introduction}
\fi
\IEEEPARstart{I}{verse} Problems (IPs) exist in various imaging techniques, such as Computed Tomography (CT) \cite{b1, b2} and Magnetic Resonance Imaging (MRI) \cite{b3, b4, b5}. The task of the inverse problem in imaging is to reconstruct an unknown image from noisy measurements. As IPs are ill-posed problems, reliable priors are needed to improve their invertibility. Conventional methods usually adopt hand-crafted priors, such as $L_1$ regularization, Total Variation ($TV$) regularization, and so forth. This type of priors has a poor discriminative ability which may cause undesired solutions. Recently, Deep Learning (DL) has offered a different paradigm for IPs due to its remarkable nonlinear fitting and feature extraction abilities. Reported work has demonstrated that DL-based methods can solve certain problems that hand-crafted prior-based algorithms cannot effectively address \cite{b6, b7}.

Like other imaging fields, researchers in Electrical Impedance Tomography (EIT) also devote themselves to IPs using DL. As a fast functional imaging modality, EIT is widely investigated in industrial processes \cite{b8, b9} and bio-medicine \cite{b10, b11, b12}. However, its further development is limited by the low spatial resolution caused by the severe ill-posedness of the EIT inversion. Therefore, DL has naturally been leveraged to solve such a challenging problem. 

\begin{figure*}[ht!]
\centerline{\includegraphics[width=160 mm]{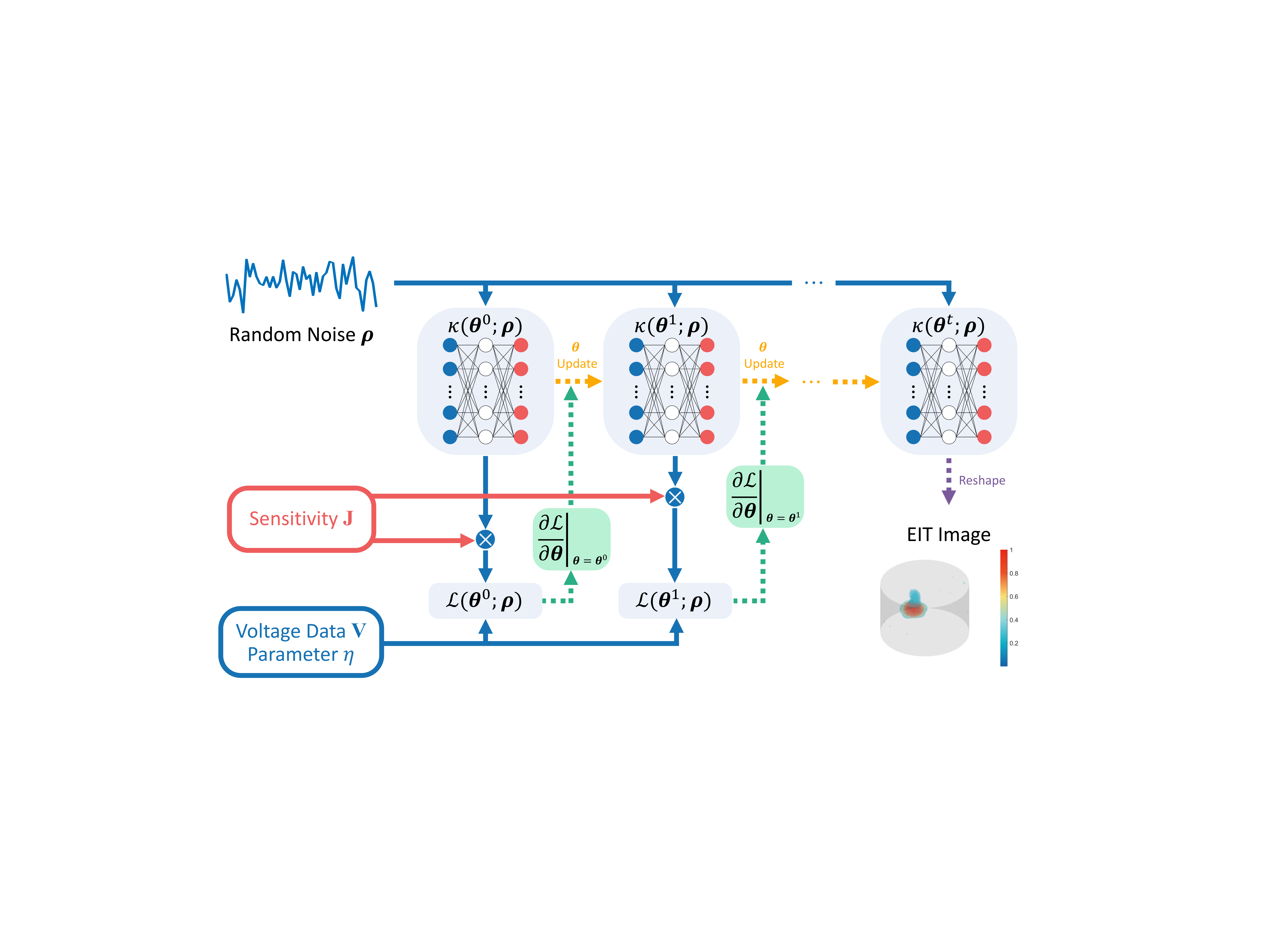}}
\caption{The flowchart of the regularized shallow image prior (R-SIP) based EIT image reconstruction algorithm. $\boldsymbol{\theta}^0$ represents the initial MLP's parameters. The cross symbol in a blue circle denotes the matrix product. Dashed arrows either denote or point toward a data operation and solid arrows represent data flow.}
\label{principle}
\end{figure*}

The past several years have witnessed many DL-based EIT image reconstruction algorithms. 
% In general, these algorithms can be divided into two categories according to whether the neural network (NN) is trained, i.e. trained neural network (TNN) based methods and untrained neural network (UNN) based methods. For the TNN-based methods, 
For example, Hamilton et al. proposed to use U-Net as a post-processing method for deconvolving the convolved direct reconstructions of the D-bar method and demonstrated structure-enhanced EIT images \cite{b13}. Wei et al. also adopted the U-Net architecture to process the neural network's multichannel inputs originating from the
dominant parts of the Induced Contrast Current (ICC) and produced quality-improved EIT images. In addition, a V-shaped dense denoising convolutional neural network was proposed to enhance the EIT images reconstructed using the model-based algorithm \cite{b14}. Wang et al. designed a Convolutional Neural Network (CNN) to post-process the results of the model-based algorithm \cite{b15}. These methods either use the reconstructed images or certain intermediate quantities of other model-based algorithms as the neural network's inputs. Another type of DL-based approach directly learns the mapping from the voltage measurements to the conductivity distribution. For instance, a CNN was proposed to directly solve the inverse problem of EIT \cite{b16}. In \cite{b17}, a densely connected U-Net was adopted to solve the same problem. To endow DL-based algorithms with interpretability, model-based learning is explored. Herzberg et al. proposed a Graph Convolutional Newton-type Method (GCNM) for solving EIT image reconstruction and their results illustrated a good generalization ability on distinct domain shapes and meshes \cite{b18}. Colibazzi et al. designed an unrolled Gauss-Newton network for EIT inversion and demonstrated improved EIT image quality compared with the model-based Gauss-Newton algorithm \cite{b19}. Other than these single-modal algorithms, DL-based multi-modal methods have been reported that utilize information from auxiliary imaging modalities to further improve EIT image quality \cite{b20, b21}. 

The aforementioned approaches rely on Trained Neural Networks (TNNs), which usually need a significant amount of training data. This requirement can limit the algorithm performance and generalization capability, as it depends heavily on the data volume and quality.
% Although TNN-based methods demonstrated noteworthy advancements in the improvement of EIT image quality. These methods require a large amount of data for training and the performance of the TNN-based methods is limited by the properties of the training data. 
UNNP-based methods can address the data-dependency problem and generate comparable results to TNN-based approaches. The initial UNNP-based work was described in \cite{b22}, which established a new connection between the inverse problem and DL. Within the UNNP framework, a neural network acts as the regularizer to the inverse problem. Subsequent research has proposed various UNNP-based approaches \cite{b23, b24, b25, b26}. In the context of EIT, Liu et al. introduced the UNNP-based method for 2D reconstruction and showed superior image quality compared to conventional model-based algorithms \cite{b27}.

The UNNP-based algorithm has demonstrated great potential in EIT inversion and is worth further exploration. Existing UNNP-based approaches usually involve carefully designed deep architectures as a deep image prior. In this paper, we explore the possibility of employing 3-layer MLPs in EIT image reconstruction in an untrained manner. Additionally, we introduce hand-crafted smoothing regularization (e.g. $TV$ or $Laplacian$ regularization) to improve image quality. We refer to this type of priors as Regularized Shallow Image Priors (R-SIPs). The contributions of this study are as follows:
\begin{itemize}
\item We prove that a 3-layer MLP regularizer with a hand-crafted prior, i.e. R-SIP, is an effective regularization strategy for 2D and 3D EIT imaging. R-SIP-based algorithms demonstrate comparable performance to DIP-based algorithms and outperform to conventional regularization-based algorithms.
\item The architecture of MLPs can be chosen from a large range, which minimizes the dependence on the neural network architecture and provides an easy way to leverage the advantages of UNNPs.  
\end{itemize}

The rest of this paper is organized as follows:
Section II introduces the principle of EIT. Section III details the proposed image reconstruction algorithm. Section IV describes the experimental setup. Section V gives the simulation and experimental results and analyzes the properties of our method. Section VI draws the conclusion and discusses future work.

\section{EIT Image Reconstruction}
Time-difference imaging is adopted in this study. It is based on the linearized EIT forward model \cite{b28}:
\begin{align}\label{linear_model}
\mathbf{V} = \mathbf{J} \boldsymbol{\sigma}, 
\end{align}
where $\mathbf{V}=(\mathbf{V}_o - \mathbf{V}_r)/\mathbf{V}_r \in \mathbb{R}^{M}$ and $\boldsymbol{\sigma} = -(\boldsymbol{\sigma}_o - \boldsymbol{\sigma}_r)/\boldsymbol{\sigma}_r \in \mathbb{R}^{N}$ represent normalized voltage measurements and conductivity distribution, respectively. $\mathbf{V}_o \in \mathbb{R}^{M}$ denotes the voltage measurements at the observation time point and $ \mathbf{V}_r \in \mathbb{R}^{M}$ stands for those at the reference time point. Similarly, $\boldsymbol{\sigma}_o \in \mathbb{R}^{N}$ represents the conductivity distribution at the observation time point and $\boldsymbol{\sigma}_r \in \mathbb{R}^{N}$ denotes those at the reference time point. $M$ and $N$ account for the number of measurements and image pixels/voxels, respectively. Vector division $'/'$ means element-wise division. $\mathbf{J} \in \mathbb{R}^{M\times N}$ represents the normalized sensitivity matrix.
% which equals to $\boldsymbol{\mathcal{J}} ./ \left(\boldsymbol{\mathcal{J}} \mathbf{1}_N\right)$. The symbol '$./$' means each element of $\mathcal{J}$ in a row is divided by the element of $\left(\boldsymbol{\mathcal{J}} \mathbf{1}_N\right)$ at that row.

Due to the ill-posedness of EIT inversion, priors are required to reduce the feasible set size. Thus, EIT image reconstruction can be generally expressed as:
\begin{equation}
\min_{\boldsymbol{\sigma}} \quad  ||\mathbf{V} - \mathbf{J} \boldsymbol{\sigma} ||^2 + \phi  R(\boldsymbol{\sigma}),
\end{equation}
where $||\cdot||$ denotes the $l_2$ norm. $R:~\mathbb{R}^{n} \rightarrow \mathbb{R}$ stands for the regularization function and $\phi > 0$ is the parameter.

\section{Methodology}
This section first introduces the concept of image representation with MLPs. Then we describe the Regularized Shallow Image Prior (R-SIP) based 3D image reconstruction algorithm. Finally, we introduce its extension to the 2D case.

\subsection{Image Representation with MLP}
In Kernel method \cite{kernel}, the unknown conductivity distribution $\boldsymbol{\sigma}$ can be represented by a linear equation:
\begin{equation}\label{kerneleq}
\boldsymbol{\sigma} = \mathbf{K} \boldsymbol{\tau},
\end{equation}
where $\mathbf{K} \in \mathbb{R}^{N \times N}$ is the kernel matrix that encodes a certain prior and $\boldsymbol{\tau} \in \mathbb{R}^{N}$ denotes the kernel coefficients.  From the network viewpoint, the Kernel method can be regarded as an iterative two-layer neural network-based reconstruction algorithm \cite{dip1}. Extending this idea, like \cite{dip1, dip2, dip3} does, $\boldsymbol{\sigma}$ can be generally expressed by the nonlinear representation:
\begin{equation}
\boldsymbol{\sigma} = \kappa(\boldsymbol{\theta}; \boldsymbol{\rho}),
\end{equation}
where $\kappa$ is a neural network. In this study, we refer to $\kappa$ as an MLP. $\boldsymbol{\theta} \in \mathbb{R}^Q$ denotes the parameters of the MLP and $Q$ is the number of the parameters. $\boldsymbol{\rho} \in \mathbb{R}^G$ stands for the random noise, which is the input of the MLP and $G$ represents the number of input neurons.

\subsection{Regularized Shallow Image Prior for 3D EIT Image Reconstruction}
We first consider the 3D EIT image reconstruction problem, which can be formulated as a penalized nonlinear optimization problem:
\begin{align} \label{OP}
\min_{\boldsymbol{\theta}} \quad  \mathcal{L}(\boldsymbol{\theta}; \boldsymbol{\rho}))  =  ||\mathbf{V} - \mathbf{J} \kappa(\boldsymbol{\theta}; \boldsymbol{\rho})||^2 + \frac{\eta}{N}  R_{H}(\kappa(\boldsymbol{\theta}; \boldsymbol{\rho})),
\end{align}
where $R_{H}$ denotes the hand-crafted regularization function. $\eta > 0$ represents the parameter for $R_{H}$. In the later section (Section 5), we demonstrate that a natural image can not always be guaranteed with only SIP. Therefore, hand-crafted priors are added to further regularize the image. For simplicity, we denote $\kappa(\boldsymbol{\theta}; \boldsymbol{\rho})$ as $\boldsymbol{\sigma}$ thereafter.

We investigate the performance of R-SIP with two hand-crafted smoothing regularizations, i.e. Total Variation ($TV$) and $Laplacian$ regularizations. For $TV$ regularization, $R_H = R_{TV}$, where $R_{TV}$ denotes the isotropic $TV$ regularization function, which provides global prior information and facilitates edge-preserving and noise smoothing. The expression of $R_{TV}$ is \cite{TV1}:
\begin{align}
R_{TV}(\boldsymbol{\sigma}) 
 & = \sum_{i, j, k} \sqrt{
\begin{aligned}
& ((\nabla_h \boldsymbol{\sigma})_{i,j,k})^2 +((\nabla_w \boldsymbol{\sigma})_{i,j, k})^2  \\ & + ((\nabla_d \boldsymbol{\sigma})_{i,j, k})^2 +\epsilon 
\end{aligned}} ~, \\
(\nabla_h \boldsymbol{\sigma})_{i,j,k}  &= (\boldsymbol{\sigma})_{i+1,j,k} - (\boldsymbol{\sigma})_{i,j,k} ~,  \\
(\nabla_w \boldsymbol{\sigma})_{i,j,k}  &= (\boldsymbol{\sigma})_{i,j+1, k} - (\boldsymbol{\sigma})_{i,j,k} ~,  \\
(\nabla_d \boldsymbol{\sigma})_{i,j,k}  &= (\boldsymbol{\sigma})_{i,j,k+1} - (\boldsymbol{\sigma})_{i,j,k} ~,
\end{align}
where $\nabla_h \boldsymbol{\sigma} \in \mathbb{R}^{N}$, $\nabla_w \boldsymbol{\sigma} \in \mathbb{R}^{N}$ and $\nabla_d \boldsymbol{\sigma} \in \mathbb{R}^{N}$ represent the gradient components of $\boldsymbol{\sigma}$ along the height, width, and depth directions, respectively. $i$, $j$ and $k$ represent the voxel indices when $\boldsymbol{\sigma}$ is reshaped to a 3D image. $\epsilon$ is a small constant whose value is set as $10^{-10}$ throughout this paper.

For $Laplacian$ regularization, $R_H = R_{Lap}$, where $R_{Lap}$ represents the $Laplacian$ regularization function, which is a local prior and smooths sudden intensity variation by punishing its second-order gradient. $R_{Lap}$ is expressed by \cite{b28}:
\begin{equation}
R_{Lap}(\boldsymbol{\sigma}) = ||\mathbf{L} \boldsymbol{\sigma}||^2,
\end{equation}
where $\mathbf{L} \in \mathbb{R}^{N \times N}$ is the $Laplacian$ matrix and $\mathbf{L} \boldsymbol{\sigma}$ is the discrete $Laplacian$. $||\mathbf{L} \boldsymbol{\sigma}||^2$ can be acquired by convolving the 3D version of $\boldsymbol{\sigma}$ with a small kernel $\boldsymbol{\varsigma} \in \mathbb{R}^{3 \times 3}$ , then adding up its all squared elements. The kernel adopted in this study is:

% \begin{tikzpicture}[every node/.style={anchor=north east,fill=white,minimum width=1 mm,minimum height=1 mm}]
% \matrix (mA) [draw,matrix of math nodes]
% {
% (22,1,3) & (1,1,3) & (1,1,3) \\
% (1,1,3) & (1,1,3) & (1,1,3) \\
% (1,1,3) & (1,1,3) & (1,1,3)  \\
% (1,1,3) & (1,1,3) & (1,1,3)  \\
% };

% \matrix (mB) [draw,matrix of math nodes] at ($(mA.south west)+(3,-0.1)$)
% {
% (43,1,3) & (1,1,3) & (1,1,3)  \\
% (1,1,3) & (1,1,3) & (1,1,3)  \\
% (1,1,3) & (1,1,3) & (1,1,3)  \\
% (1,1,3) & (1,1,3) & (1,1,3)  \\
% };

% \matrix (mC) [draw,matrix of math nodes] at ($(mB.south west)+(3,-0.1)$)
% {
% (1,1,3) & (1,1,3) & (1,1,3)  \\
% (1,1,3) & (1,1,3) & (1,1,3)  \\
% (1,1,3) & (1,1,3) & (1,1,3)  \\
% (1,1,3) & (1,1,3) & (1,1,3)  \\
% };

% \draw[dashed](mA.north east)--(mC.north east);
% \draw[dashed](mA.north west)--(mC.north west);
% \draw[dashed](mA.south east)--(mC.south east);
% \end{tikzpicture}

\begin{equation}
\boldsymbol{\varsigma} =
\frac{1}{26}
\begin{bmatrix}
\begin{bmatrix}
    2  &  3  &  2  \\
    3  &  6  &  3  \\
    2  &  3  &  2 
\end{bmatrix};
\begin{bmatrix}
    3  &  6  &  3  \\
    6  & -88 &  6  \\
    3  &  6  &  3 
\end{bmatrix};
\begin{bmatrix}
    2  &  3  &  2  \\
    3  &  6  &  3  \\
    2  &  3  &  2 
\end{bmatrix}
\end{bmatrix}.
\end{equation}

The problem in (\ref{OP}) can be solved with various standard optimization algorithms. In this study, we solve (\ref{OP}) with Adam \cite{AdamAlgorithm}, a widely used optimizer in network training. In order to perform Adam, at each iteration, the first step is to calculate the gradient of $\mathcal{L}$ with respect to neural network parameters $\boldsymbol{\theta}$, i.e.:
\begin{align} \label{grad}
\frac{\partial \mathcal{L}}{\partial \boldsymbol{\theta}} = & \left(\frac{\partial F}{\partial \boldsymbol{\sigma}} \right)^T \frac{\partial \boldsymbol{\sigma}}{\partial \boldsymbol{\theta}} + \frac{\eta}{N} \left(\frac{\partial R_{H}}{\partial \boldsymbol{\sigma}} \right)^T \frac{\partial \boldsymbol{\sigma}}{\partial \boldsymbol{\theta}},
\end{align}
where we denote $||\mathbf{V} - \mathbf{J} \boldsymbol{\sigma}||^2$ as $F$. As each term in (\ref{OP}) is expressed by $Tensors$, the PyTorch built-in class, we implement (\ref{grad}) by using PyTorch's automatic differentiation engine called $torch.autograd$. Afterward, the neural network parameters can be updated according to the parameter update rules of Adam. The number of iterations $t$ is considered as the algorithm' parameter. Thus, there are three parameters for R-SIP-based reconstruction, i.e $\eta$, $t$, and the learning rate of Adam (denoted by $lr$). The flowchart of the R-SIP algorithm is visually illustrated in Fig. \ref{principle}.

% The stopping criteria are defined by two conditions. The first one is the maximum iteration $t_{\textrm{max}} \in \mathbb{Z}^+$ and the second one is the tolerance $\vartheta >0$ which is defined as:
% \begin{equation} 
% \frac{||\mathcal{L}^{t+1}-\mathcal{L}^{t}||}{||\mathcal{L}^{t}||}<\vartheta.
% \end{equation}

% If any of the conditions are satisfied, OGLL will stop. 
% The implementation of OGLL is summarised in $\mathbf{Algorithm ~1}$. 

\subsection{Extension to 2D EIT Image Reconstruction}
The 3D image reconstruction algorithm can be directly applied to the 2D case with two modifications. First, the $TV$ regularization only involves height and width directions. Second, the $Laplacian$ matrix used in the 2D situation is the same as that in \cite{b28}. The other aspects of the algorithm remain the same.
% If not specified, we fix the iteration number to 2000 for both simulation and real experiment studies. The learning rate of Adam is set as $10^{-4}$ and $5 \times 10^{-4}$ for 2D and 3D reconstruction based on simulation data, and is set as $10^{-4}$ for real experiments. Only the regularization parameter requires tuning. To avoid confusion, we denote the $\zeta$ as $\lambda$ and $\mu$ for the parameters of $TV$ and $Laplacian$ regularization, respectively.

\section{Experimental Setup}
\subsection{Data for Evaluation}
\subsubsection{2D Simulation}
\begin{figure}[!t]
\centerline{\includegraphics[width=85 mm]{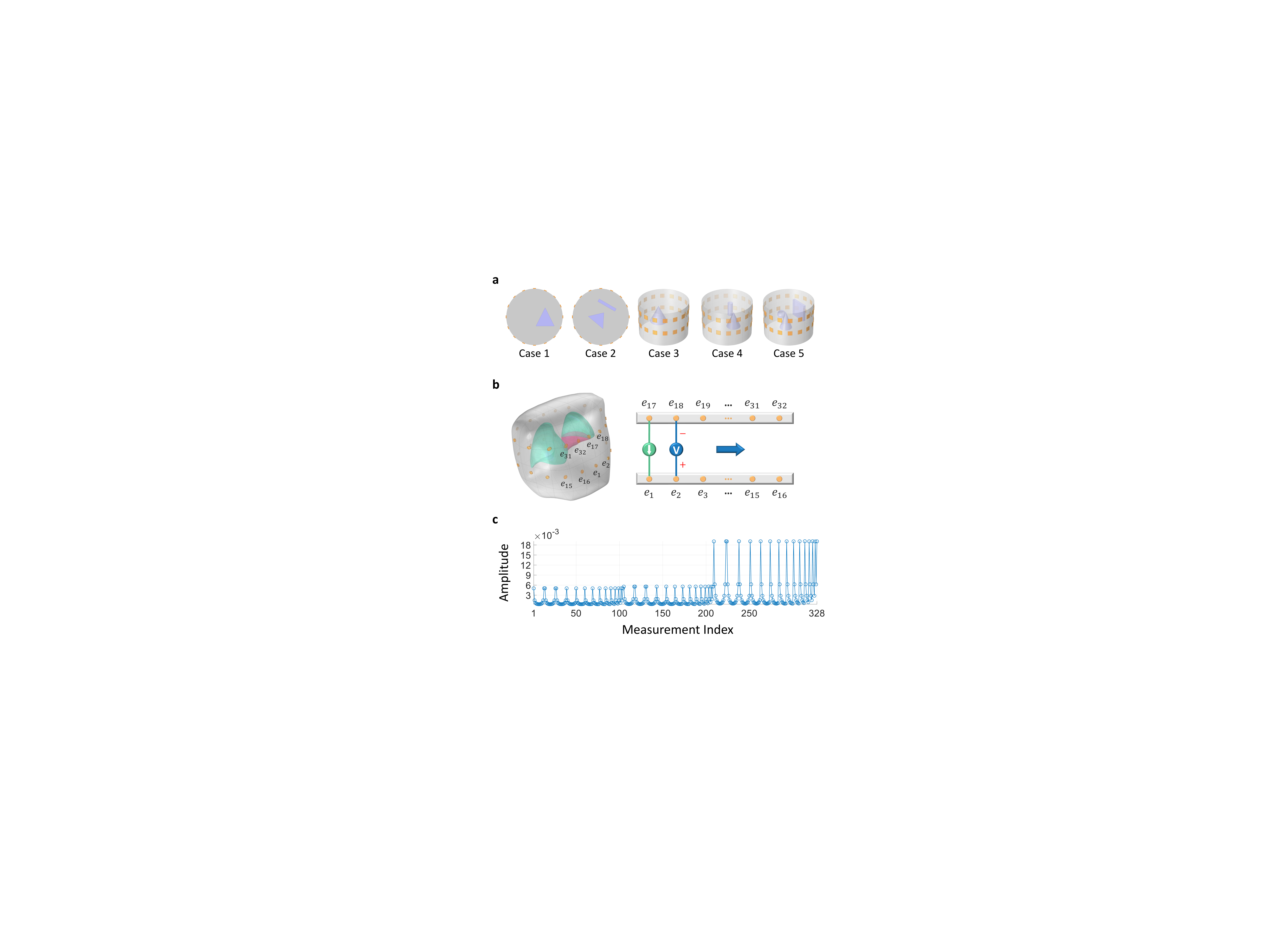}}
\caption{\textbf{a}, Simulated 2D and 3D phantoms. The inclusions are highlighted with bluish violet. \textbf{b}, Thoracic phantom with the illustration of the cross-layer stimulation-measurement pattern. In \textbf{a} and \textbf{b}, the golden color indicates electrodes. \textbf{c}, The measured reference voltage data.}
\label{phan}
\end{figure}

We modeled a circular region and attached 16 electrodes to its boundary (see Fig. \ref{phan}a) in COMSOL Multiphysics. In this circular region, two types of conductivity distributions (labeled as case 1 and case 2) are constructed. Case 1 simulates a triangular object whose conductivity is 0.8 $S/m$. Case 2 includes a triangular object with a conductivity of 0.4 $S/m$ and a rectangular bar with a conductivity of 1.2 $S/m$. The background conductivity is 2 $S/m$ for cases 1 and 2. The adjacent measurement protocol \cite{b29} is adopted for 2D imaging, producing 104 voltage measurements.

\begin{figure}[!t]
\centerline{\includegraphics[width=88 mm]{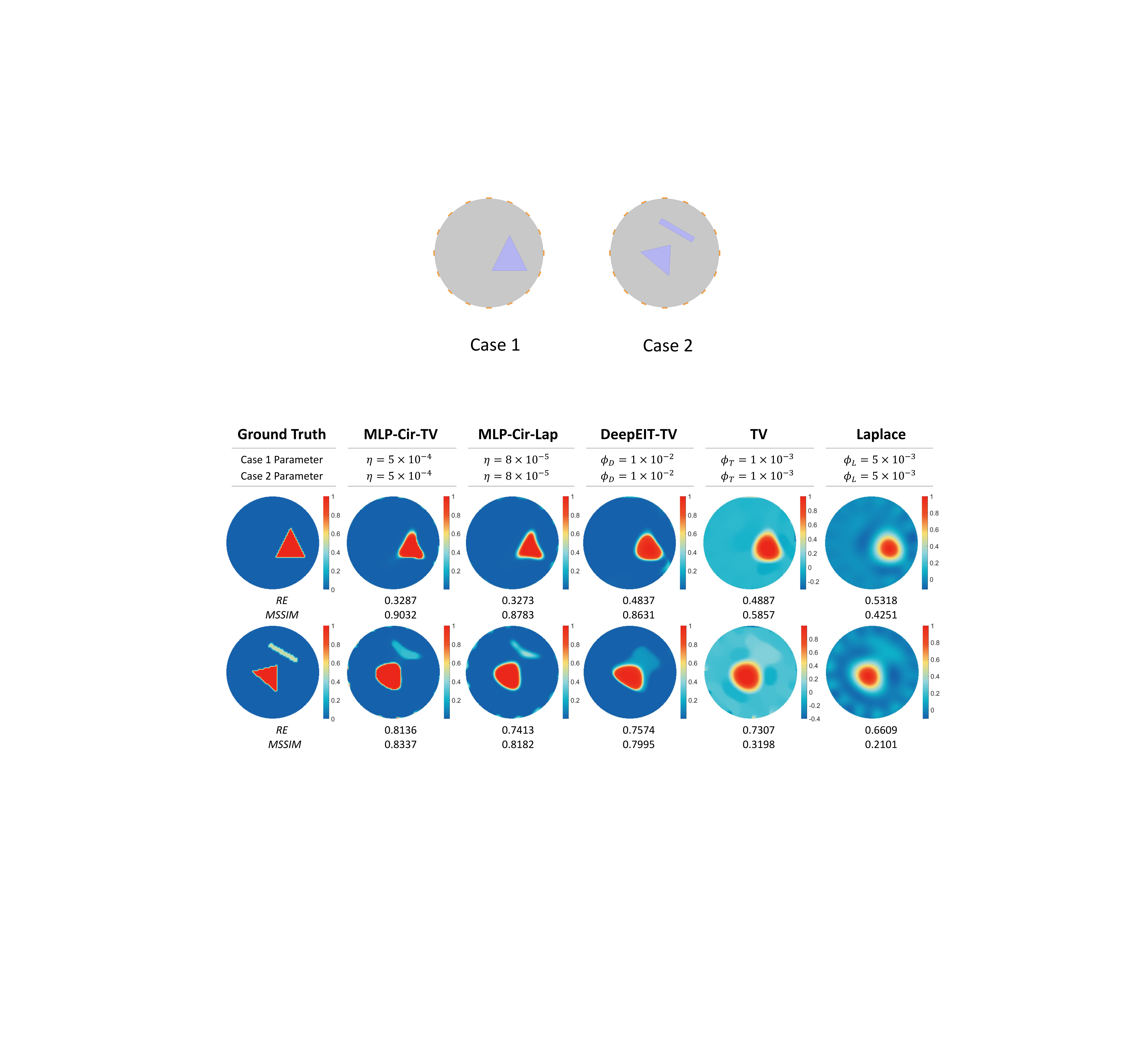}}
\caption{Image reconstruction comparison on simulated 2D EIT data.}
\label{2DResult}
\end{figure}

\begin{figure*}[!t]
\centerline{\includegraphics[width=182 mm]{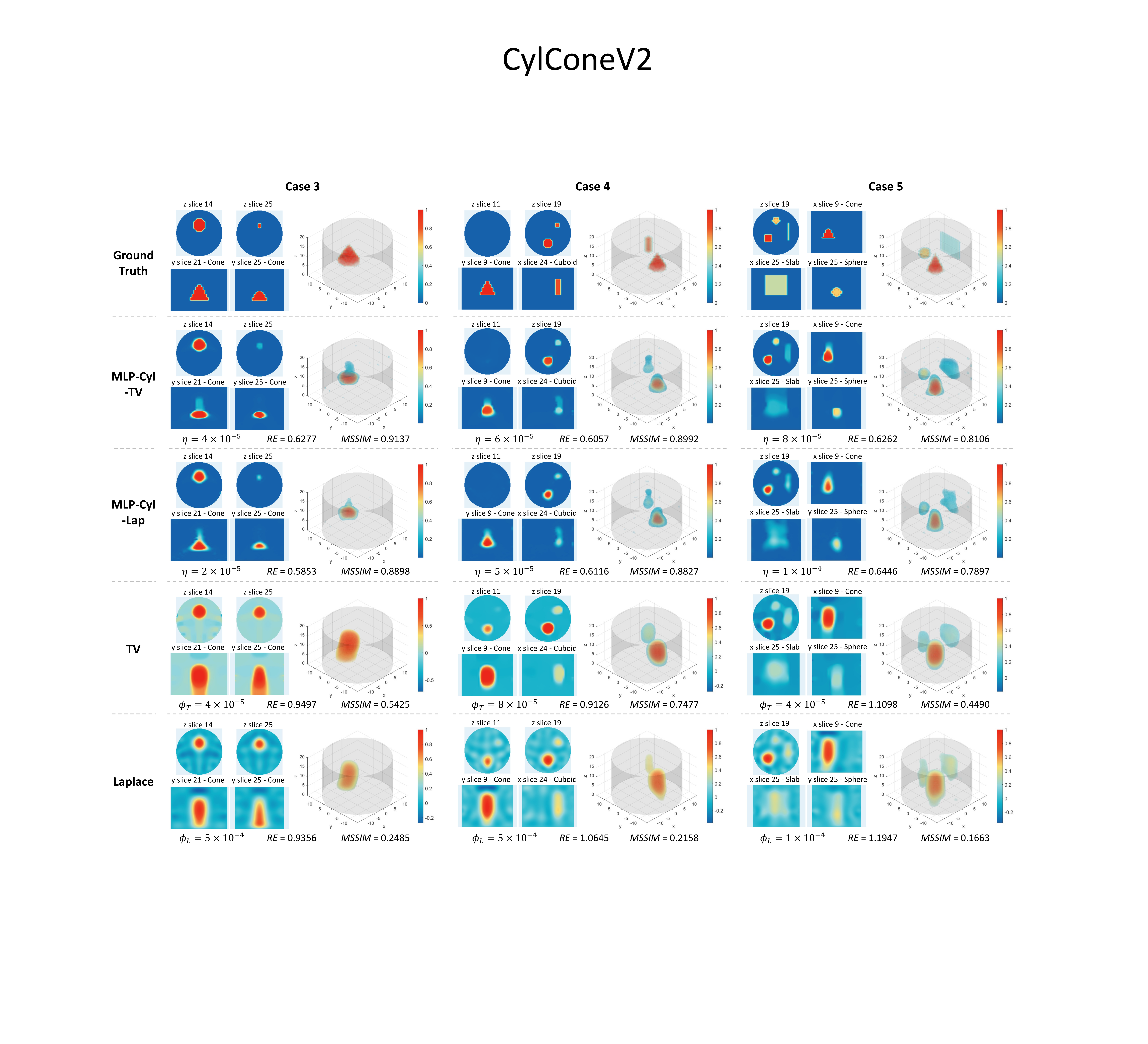}}
\caption{Image reconstruction comparison on simulated 3D EIT data. The smaller x/y/z slice number corresponds to the smaller x/y/z coordinate and vice versa. The word following the slice number is the inclusion's name. As the z slice contains all inclusions, we do not add any suffixes.}
\label{3DCylResult}
\end{figure*}

\subsubsection{3D Simulation}
For 3D imaging, we modeled cylindrical and thoracic regions for EIT imaging (see Fig. \ref{phan}a and \ref{phan}b). Thirty-two electrodes arranged as 2 layers are placed on the boundaries, and each electrode layer includes 16 electrodes. For the cylindrical region, we utilize various objects to form three types of conductivity distributions (labeled as cases 3, 4, and 5). For case 3, we put a cone in the imaging region and set its conductivity as 1 $S/m$; For case 4, we place a cone and a cylinder in the imaging region. The conductivities of the cone and cylinder are 0.8 $S/m$ and 1.2 $S/m$, respectively. For case 5, we put a cone, a sphere, and a cuboidal slab in the imaging region, and their conductivities are 0.4 $S/m$, 1 $S/m$, and 1.2 $S/m$ accordingly. The background medium's conductivity is 2 $S/m$ for all cases.
For the thoracic region, we modeled lungs with pulmonary edema. The background conductivity is 0.24 $S/m$. The conductivity of healthy lung tissue (indicated by green) and unhealthy lung tissue (indicated by fuchsia) are 0.06 $S/m$ and 0.24 $S/m$, respectively.

\begin{figure*}[!t]
\centerline{\includegraphics[width=170 mm]{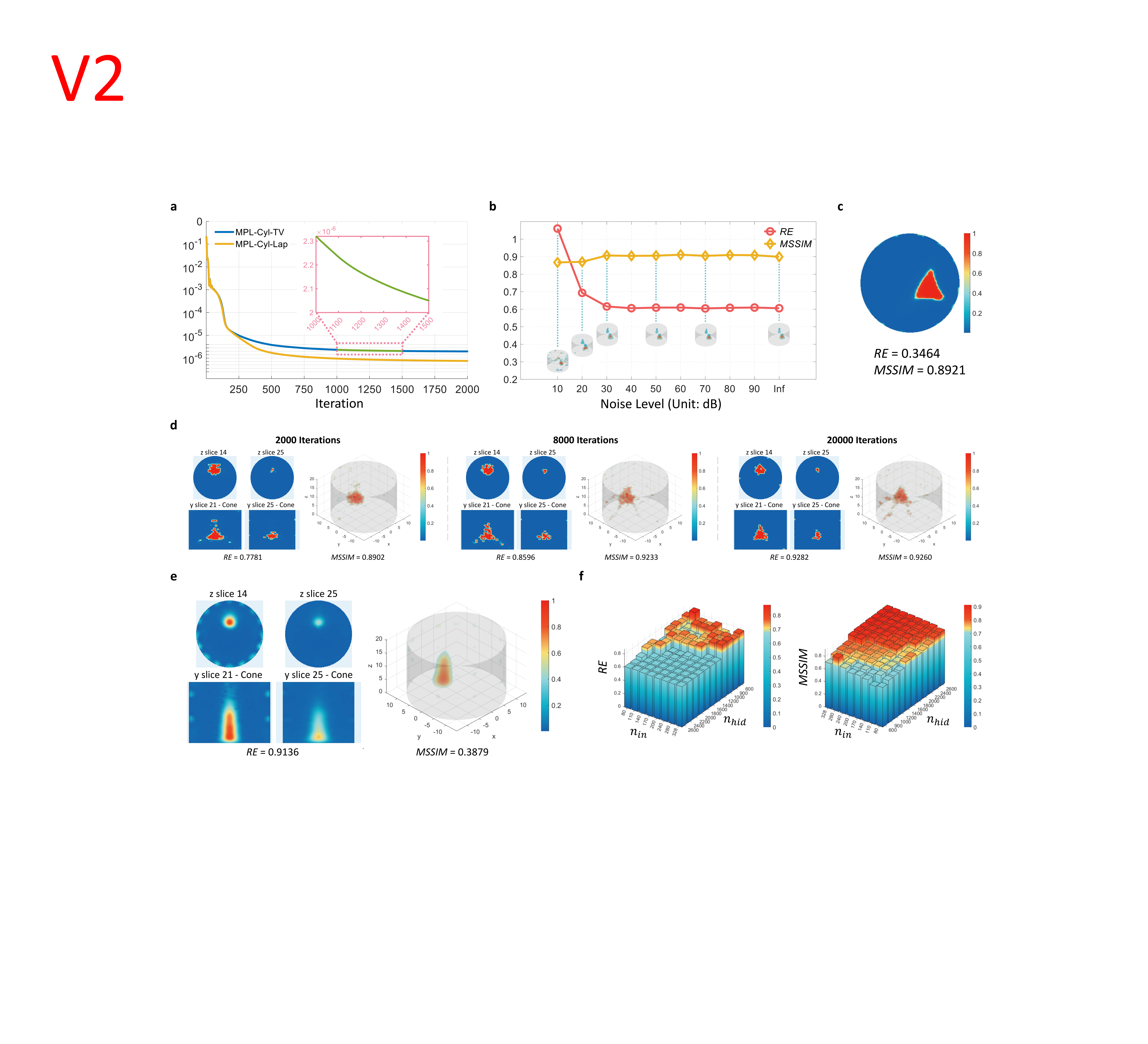}}
\caption{\textbf{a}, Convergence of the proposed algorithm. \textbf{b}, The noise resistance performance of our method. \textbf{c}, The image reconstruction result of case 1 without hand-crafted prior. \textbf{d}, Image reconstruction results of case 3 without hand-crafted prior. \textbf{e}, Image reconstruction results of case 3 without voltage data from the cross-layer stimulation-measurement pattern. \textbf{f}, $RE$ and $MSSIM$ variation with the different numbers of MLP's input and hidden neurons.} 
\label{properties}
\end{figure*}

We number the electrodes ($e_1$, $e_2$, ..., $e_{32}$) in the thoracic model to describe our customized sensing strategy for 3D imaging (see Fig. \ref{phan}b). Our sensing strategy is divided into 3 stages, including 2 intra-layer stimulation-measurement patterns and 1 cross-layer stimulation-measurement pattern. The adjacent strategy \cite{b29} is adopted for intra-layer stimulation and measurement. Following the adjacent strategy, we first collect voltages at the bottom layer of electrodes and then collect voltages at the top layer of electrodes. There are 104 measurements for each intra-layer stimulation-measurement pattern. 

\begin{figure*}[!t]
\centerline{\includegraphics[width=180 mm]{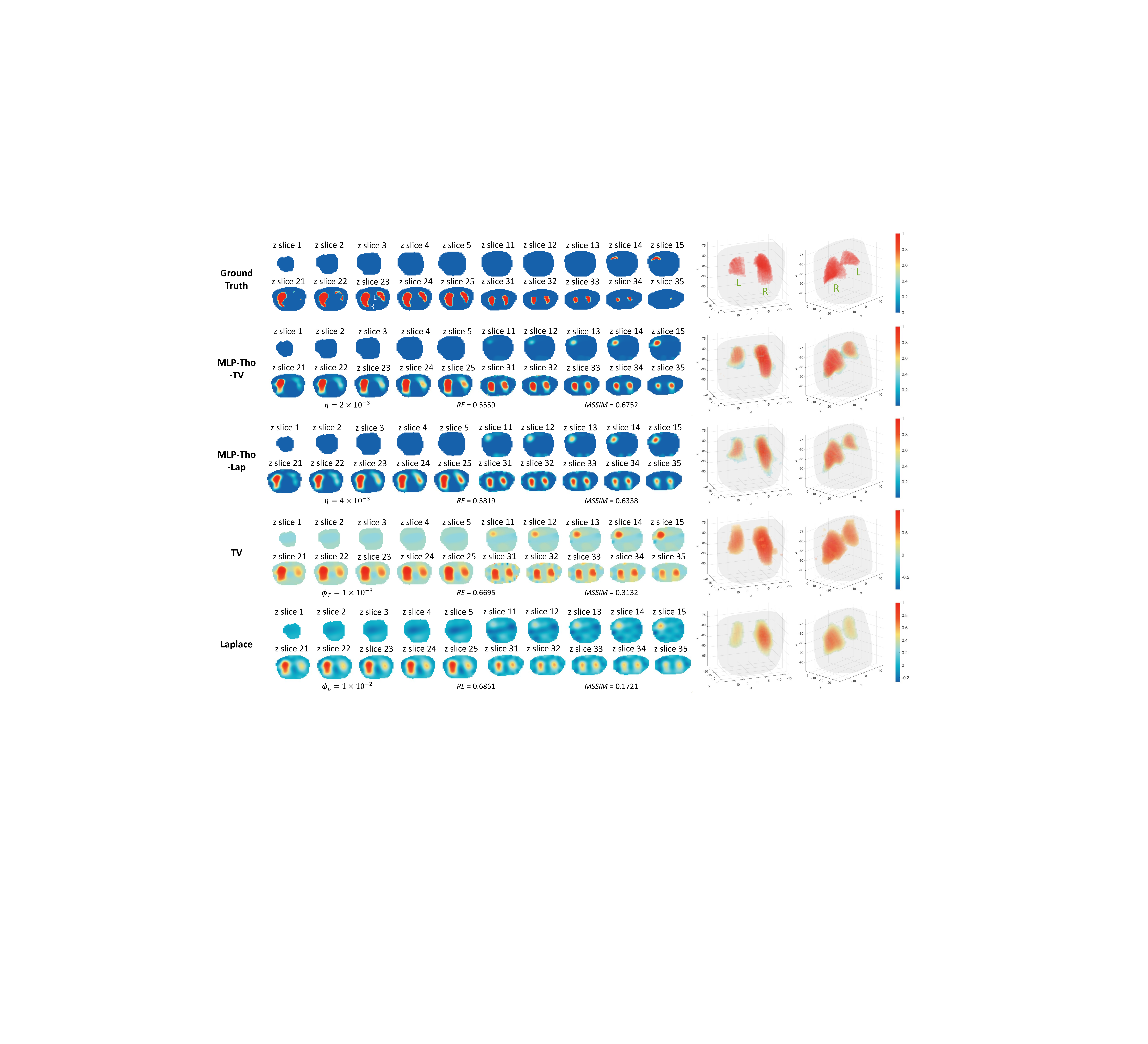}}
\caption{Thoracic image reconstruction comparison. The smaller z slice number corresponds to the smaller z coordinate and vice versa. Letters 'L' and 'R' mark the left lung and the right lung, respectively. }
\label{LungResults}
\end{figure*}

For the cross-layer stimulation-measurement pattern, the current is injected into a certain electrode at the bottom layer and flows out from the electrode right over that electrode. Then, voltages are collected from the rest electrode pairs. Each measurement pair of electrodes aligns vertically. In addition, we only collect independent voltages according to the reciprocal theorem \cite{b30}. According to this rule, the first stimulation pair is ($e_1$, $e_{17}$), and measurement pairs are ($e_2$, $e_{18}$), ($e_3$, $e_{19}$), ..., ($e_{16}$, $e_{32}$); the second stimulation pair is ($e_2$, $e_{18}$), and measurement pairs are ($e_3$, $e_{19}$), ($e_4$, $e_{20}$), ..., ($e_{16}$, $e_{32}$); ...; the last stimulation pair is ($e_{15}$, $e_{31}$), and the measurement pair is ($e_{16}$, $e_{32}$). Thus, the number of measurements is 120 for the cross-layer stimulation-measurement pattern. Eventually, we acquire 328 measurements in total for 3D imaging. The simulated reference voltage data is illustrated in Fig. \ref{phan}c.

\subsubsection{Real-World Data}
In real experiments, we imaged the cylindrical region using the EIT system \cite{eitsys} developed at The University of Edinburgh. The cylindrical EIT sensor has the same configuration as that in the simulation. For all experiments, the stimulation current frequency was 10 kHz, and the customized sensing strategy described was adopted. We place 3 sets of objects in the imaging region (see the first column of Fig. \ref{ExpResult}). The first set includes a red cuboid made of plastics, the second set contains two lung models made of silicone, and the third set consists of three objects (i.e. cone, cuboid, and a slab) made of plastics. 

\subsection{Imaging Domain Discretization}
%As EIT image reconstruction aims to recover the discrete conductivity distribution, it is essential to discretize the imaging domain before performing image reconstruction. 

In 2D imaging, we use square inverse mesh. The x/y axis between the minimum and maximum x/y coordinates of the imaging region is split into 64 segments. Thus, there are 3228 pixels within the circular region. In 3D imaging, we adopt cuboidal inverse mesh and take a similar splitting method to the 2D case. The x, y, and z axes are split into 32, 32, and 40 segments. Eventually, there are 32480 voxels in the cylindrical imaging region and 27418 voxels in the thoracic imaging region.

\subsection{Structure of MLP}
We design three decoder-type 3-layer MLPs labeled as MLP-Cir, MLP-Cyl, and MLP-Tho for imaging circular, cylindrical, and thoracic regions. Due to distinct numbers of voxels/pixels corresponding to different imaging domains, the numbers of MLP's output neurons are 3228, 32480, and 27418 for MLP-Cir, MLP-Cyl, and MLP-Tho, respectively. These 3 MLPs adopt the same input (328 neurons) and hidden (2000 neurons) layers. All MLPs take the leaky rectified linear unit $LeakyReLu$ as the activation function for the hidden layer. In biomedical applications, such as thoracic imaging and cell monitoring, the conductivity of the region of interest (RoI) usually decreases during certain physiological or pathological processes, causing $\boldsymbol{\sigma}$ within the range of $(0,~1)$. Therefore, the $Sigmoid$ is selected as the activation function of the output layer for all MLPs. If the $TV$ regularization is selected as the hand-crafted prior with an MLP, e.g. MLP-Cyl, in the R-SIP algorithm, the algorithm is then labeled as MLP-Cyl-TV. If the $Laplacian$ regularization is chosen, the algorithm is labeled as MLP-Cyl-Lap. Other MLPs also follow this labeling rule. 

\subsection{Comparison Algorithms}
We choose model-based $TV$ and $Laplacian$ regularization for comparison in 2D and 3D imaging. These two algorithms are labeled as TV and Laplace, respectively. They adopt the same expressions of $TV$ and $Laplacian$ regularization functions as those in our proposed method, and we also use Adam to solve the problem. The parameters of both TV and Laplace include the regularization parameter, the learning rate of Adam, and the number of iterations.
In the 2D comparison, we additionally choose the $TV$ regularization-based algorithm in \cite{b27}, which is a UNNP-based algorithm with a deep net. The reason that we only compare with it in 2D cases is that the neural network is designed only for 2D imaging in the original paper. This algorithm is labeled as DeepEIT-TV. 
There are three differences in implementing the DeepEIT-TV algorithm. First, we adopt the algorithm framework for time-difference imaging while the original paper conducts absolute imaging. Second, we divide the $TV$ regularization term by the number of pixels. Third, we use the $64 \times 64$ inverse mesh for 2D imaging while the original paper employs the $128 \times 128$ inverse mesh. The modifications aim to make the comparison algorithms consistent with our method in as many aspects as possible for a fair comparison. 

\subsection{Parameter Settings}
Parameter selection of all algorithms is based on trial and error and follows the settings in this subsection if not specified. For the R-SIP algorithm, we fix $t$ to 2000 for both simulation and real experiment studies. $lr$ is set as $10^{-4}$ for 2D simulation and real experiments, and set as $5 \times 10^{-4}$ for 3D simulation. Regarding TV and Laplace, we fix the learning rate as $10^{-2}$  for simulation and real experiments; the number of iterations is set as 2000 for the simulation study and 1000 for real experiments. For cases 1 and 2, the number of iterations and learning rate of the DeepEIT-TV are set as 8000 and 0.005, respectively. The regularization parameter of all algorithms is shown with the reconstructed images. The regularization parameter is denoted by $\phi_T$, $\phi_L$, and $\phi_D$ for TV, Laplace, and DeepEIT-TV, respectively.

\subsection{Quantitative Metrics}
We employ Image Relative Error ($RE$) to evaluate the performance in differentiating distinct conductivity levels, and Mean Structural Similarity Index ($MSSIM$) to evaluate the algorithm's structure preservation ability. $RE$ and $MSSIM$ are defined as:
\begin{equation}
RE = \frac{||\boldsymbol{\sigma}_p-\boldsymbol{\sigma}_g||}{||\boldsymbol{\sigma}_g||}~,
\end{equation}
\begin{equation}\label{mssim}
MSSIM = \frac{1}{N} \sum \frac{\left(2\boldsymbol{\iota}_p \boldsymbol{\iota}_{g}+\chi_1\right)\left(2\boldsymbol{\delta}_{p,g}+\chi_2\right)}{\left(\boldsymbol{\iota}_{p}^{2}+\boldsymbol{\iota}_{g}^{2}+\chi_{1}\right)\left(\boldsymbol{\delta}_{p}^{2}+\boldsymbol{\delta}_{g}^{2}+\chi_{2}\right)}~,
\end{equation}
where $\boldsymbol{\sigma}_p$ and $\boldsymbol{\sigma}_g$ denote the reconstructed and true normalized conductivity distributions. $\boldsymbol{\iota}_p \in \mathbb{R}^{N}$, $\boldsymbol{\iota}_g \in \mathbb{R}^{N}$, $\boldsymbol{\delta}_p \in \mathbb{R}^{N}$, $\boldsymbol{\delta}_g \in \mathbb{R}^{N}$ and $\boldsymbol{\delta}_{p,g} \in \mathbb{R}^{N}$ denote the local means, standard deviations and cross-covariance for $\boldsymbol{\sigma}_p$ and $\boldsymbol{\sigma}_g$, respectively. $\chi_1 = (\psi_1 \gamma)^2$ and $\chi_2 = (\psi_2 \gamma)^2$ are constants in which $\psi_1$, $\psi_2$ and $\gamma$ are set as 0.01, 0.03 and 1, respectively. The standard deviation of the isotropic Gaussian function in $MSSIM$ calculation is set as 0.35. Summation in (\ref{mssim}) is over all elements of operands. Due to the intrinsic modeling error of the linear EIT forward model, absolute reconstruction cannot be expected. Therefore, all reconstruction results are normalized with $\boldsymbol{\sigma}/max(abs(\boldsymbol{\sigma}))$ before evaluation and display. $abs()$ converts each element of a vector to its absolute value. $max()$ returns the maximum element of a vector.

\section{Results and Discussion}
\subsection{Simulation Results}
Fig. \ref{2DResult} shows the image reconstruction results using noise-free 2D data. Quantitative metrics are placed under each image. The results indicate that neural network based algorithms have better performance on background artifact suppression than conventional algorithms. Especially, our method achieves the best structure preservation performance. For example, our method reconstructs the triangular object with the sharpest vertices and the straightest edges. In addition, our method successfully preserves the rectangular bar in case 2 and separates it from the triangular object. The highest $MSSIM$ of the MLP-Cir-TV and MLP-Cir-Lap also supports this conclusion. For $RE$, our method has the lowest value in case 1, while the $RE$ in case 2 is not the lowest. This result is reasonable because, in case 2, our method estimates bigger triangular inclusion, the rectangular bar is slightly deformed, and the conductivity levels are not perfectly estimated. These factors make an impact on the result.

Fig. \ref{3DCylResult} compares algorithms on noise-free data collected from the cylindrical imaging region. Quantitative metrics are displayed under each EIT image block. For each image block, we display not only the 3D volume images but also four selected slices. For 3D images, we transparentize low-absolute-value voxels based on trials for highlighting the inclusions' structure. The results illustrate that our method achieves considerable improvements in structure prevention and background artifact suppression. For example, for cases 3, 4, and 5, only our method can reconstruct the cone while other algorithms reconstruct a round-shaped rod. For other geometrical objects in these three cases, our method can still preserve their geometrical characteristics, but other algorithms cannot make it. Regarding the quantitative metrics, our method has the lowest $RE$ and highest $MSSIM$, which further proves that our method reconstructs EIT images with the best quality compared with TV regularization-based and Laplace regularization-based algorithms.

\begin{figure*}[!t]
\centerline{\includegraphics[width=180 mm]{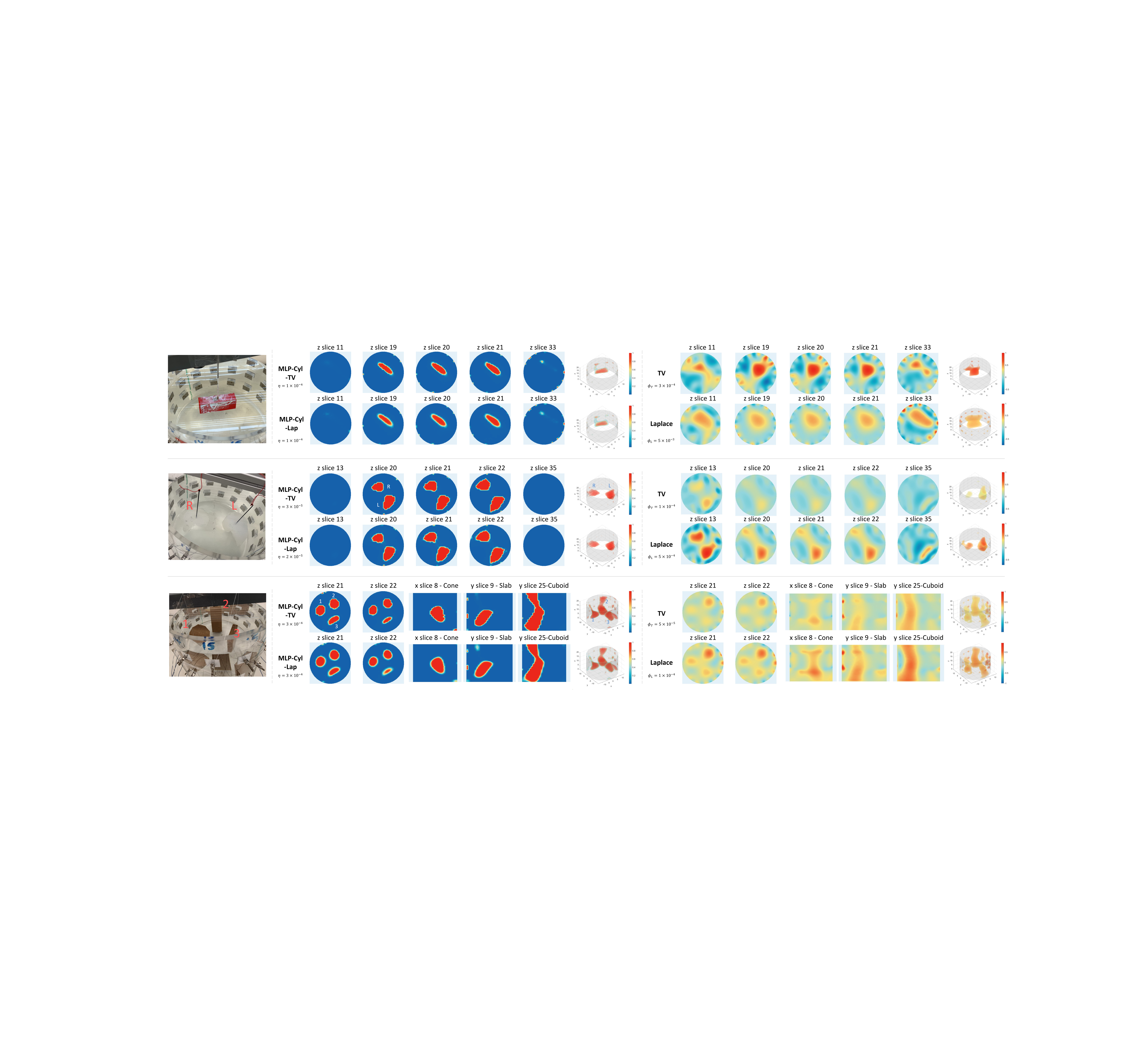}}
\caption{Image reconstruction comparison based on experimental data. The smaller z slice number corresponds to the smaller z coordinate and vice versa. For lung imaging, the left and right lungs are indicated in selected images by 'L' and 'R', respectively. Regards 3-object phantom imaging, digits '1' $\sim$ 
'3' are shown with selected images. '1' represents the cone, '2' denotes the cuboid, and '3' accounts for the slab.}
\label{ExpResult}
\end{figure*}

Fig. \ref{properties}a illustrates the convergence property of our method based on case 5. It displays the variation of the objective of (\ref{OP}) with the iteration. The vertical axis takes the logarithmic scale. The objective decreases smoothly, indicating a stable convergence property. 

We use case 4 to discuss the noise resistance ability of the proposed method. We add various levels of Gaussian noise to the voltage data, forming noisy data with SNR 10 dB $\sim$ 90 dB. The MLP-Cyl with $TV$ is chosen as the R-SIP. The variation of the $RE$ and $MSSIM$ with the SNR is illustrated in Fig. \ref{properties}b. In addition, we also display 3D images corresponding to selected noise levels. The results indicate that the performance of the proposed method becomes stable when SNR is over 30 dB. Even when the noise level is as low as 20 dB, our method can still correctly recover the inclusions, suggesting that the proposed algorithm has good noise resistance ability.

We use cases 1 and 3 as examples to discuss the function of the hand-crafted regularization, i.e. $TV$ or $Laplacian$ regularization, in our framework. 2D and 3D image reconstruction results without hand-crafted priors are shown in Fig. \ref{properties}c and in Fig. \ref{properties}d, respectively. The left image block in  Fig. \ref{properties}d adopts 2000 iterations, the middle image block takes 8000 iterations, and the right image block uses 20000 iterations. Other parameters are the same as those described in subsection 4.5. For the 2D situation, a satisfactory result is acquired. For the 3D situation, we observe that with only the neural network regularizer, our method can still suppress most background artifacts, capture the ROI and recover some structural information. However, the reconstructed images suffer from discontinuities and irregular boundaries. With the iteration increasing, this phenomenon still remains. It indicates that SIP imposes imperfect regularization on the conductivity estimation in complex setups, such as 3D imaging. Therefore, the hand-crafted prior is added to further regularize the SIP, though it is not always necessary in certain situations.

We adopt case 3 to demonstrate the necessity of the cross-layer stimulation-measurement pattern in capturing vertical structure information for 3D EIT. The MLP-Cyl with $Laplacian$ regularization is selected as the R-SIP. $lr$, $\eta$, and $t$ are set as $10^{-4}$, $5 \times 10^{-4}$ and 2000. The image reconstruction result is shown in Fig. \ref{properties}e. From the y slices 21 and 25, we see that the bottom of the inclusion reaches the bottom of the imaging region, inconsistent with the ground truth (see Fig. \ref{3DCylResult}). Thus, the cross-layer stimulation-measurement pattern can provide more vertical information in 3D imaging.  

Fig. \ref{properties}f demonstrates the influence of the number of input and hidden neurons on the algorithm performance using case 3. We do not change the number of output neurons as it is determined by the imaging region. MLP-Cyl with $Laplacian$ regularization is selected as the R-SIP. Algorithm parameters are stated in subsection 4.5. According to the results, $RE$ becomes the lowest, and $MSSIM$ reaches the highest when the number of input neurons is over 140 and the number of hidden neurons is over 1400. If the number of input neurons is small, e.g. 80, we can increase the number of hidden neurons to increase the algorithm performance. The number of optional configurations of the input layer and hidden layer is extremely large. There are two aspects that are noteworthy. First, if we fine-tune the parameters for each MLP, the performance of certain MLPs may become better. Second, if we increase the number of MLP layers, regularization may not be necessary. Although it is not possible to investigate all situations, the results offer sufficient evidence that the 3-layer MLP is an effective choice of UNNP in 2D and 3D EIT imaging. Moreover, the results suggest a wide range of options for the number of input and hidden neurons, indicating that the proposed method requires minimal effort in designing the neural network architecture.

Finally, we evaluate our method in thoracic imaging and the results are shown in Fig. \ref{LungResults}. For each algorithm, we display 3D images from two different angles and some selected slices. We can see that the proposed method demonstrates better performance on background artifact suppression and structure preservation. Especially, for the left lung, our method generates a more accurate structure compared to other methods. For example, the left lung in the z slices 21 $\sim$ 25 of the $TV$ results is an ellipse, which is far from the ground truth. In addition, our method achieves the lowest $RE$ and highest $MSSIM$, further validating its superior performance.

\subsection{Real Experiment Results}
Fig. \ref{ExpResult} gives the image reconstruction results based on experimental data. For each algorithm, we display the 3D image and selected 2D slices. From the 2D images, we can observe that our method has better performance on background artifact suppression than other algorithms. In addition, the proposed method can reconstruct objects with more precise structures than conventional algorithms. For example, z slices 19 $\sim$ 21 of the cuboid image indicate that our method can reconstruct the bar-type shape while TV and Laplace reconstruct a circular shape. For lung model imaging, our method also reconstructs the lung model with a more accurate shape.
For the three-object phantom, we can reconstruct the slab while other algorithms can not differentiate the slab from other objects (see z slices 21 and 22). The x slice 8 shows that only our approach can reconstruct the downward cone which other algorithms fail to do. It is noted that the objects in the images of the three objects imaging distort, which is possibly caused by the imperfection of the EIT measurements and inevitable errors in the experiments.

\section{Conclusion}
 We propose an image reconstruction framework based on the shallow image prior with hand-crafted regularization for 2D and 3D EIT image reconstruction. The shallow image prior is implemented by representing the conductivity distribution with a 3-layer MLP, and the hand-crafted regularization is combined to further improve image quality. Our method provides a way to develop UNNP-based algorithms, which considerably reduce the time cost of designing the neural network architecture and release the algorithm's dependency on it. We demonstrate that our method has a smooth convergence property. Simulation and real experiments show that our method achieves the best performance compared with given algorithms, especially in structure preservation. Future work will extend this method to multi-frequency EIT imaging and apply it to quantitative analysis in tissue engineering.

\end{document}